\documentclass[12pt]{article}
\usepackage{graphics}
\usepackage{amssymb}

\newcommand{\C}{\mathbb{C}}
\newcommand{\R}{\mathbb{R}}
\newcommand{\e}{\mathrm{e}}
\newcommand{\GG}{\mathcal{G}}
\newcommand{\HH}{\mathcal{H}}
\newcommand{\JJ}{\mathcal{J}}
\newcommand{\OO}{\mathcal{O}}
\begin{document}

\title{An approximation to $\delta'$ couplings on graphs}
\author{T.~Cheon$^{a}$ and P.~Exner$^{b,c}$}
\date{}
\maketitle
\begin{quote}
{\small \em a) Laboratory of Physics, Kochi University of Technology, \\
\phantom{e)x}Tosa Yamada, Kochi 782-8502, Japan\\
 b) Department of Theoretical Physics, Nuclear
Physics Institute, \\ \phantom{e)x}Academy of Sciences, 25068, \v
Re\v z, Czech Republic \\
 c) Doppler Institute, Czech Technical University, B\v{r}ehov{\'a} 7,\\
\phantom{e)x}11519 Prague, Czech Republic \\
 \rm \phantom{e)x}taksu.cheon@kochi-tex.ac.jp, exner@ujf.cas.cz}
\vspace{8mm}

\noindent {\small We discuss a general parametrization for
vertices of quantum graphs and show, in particular, how the
$\delta'_s$ and $\delta'$ coupling at an $n$ edge vertex can be
approximated by means of $n+1$ couplings of the $\delta$ type
provided the latter are properly scaled.}
\end{quote}

%%%%%%%%%%%%%%%%%%%%%%%%%%%%%%%%%%%%%%%%%%%%%%%%%%%%%%%%%%%%%%%%%%%

\noindent Quantum graphs became in the last decade a useful and
versatile tool to describe several classes of physical systems, in
particular, various combinations of quantum wires. There are
numerous papers devoted to the subject and we restrict ourselves
to mentioning the bibliography given in \cite{KS99, Ku04}, where
also basic concepts of theory are discussed.

The purpose of this letter is twofold. First of all we want to
draw attention to useful parametrization of a general coupling at
graph vertices whose advantages in the present context remained so
far unnoticed. Second and more important, we address the question
of physical meaning of such a coupling and suggest an answer
illustrating it on a pair of simple nontrivial examples of the
so-called $\delta'_s$ and $\delta'$ couplings \cite{Ex95, Ex96a}.

We consider a free spinless particle on a graph, with the
Hamiltonian which acts as $H\psi_j= -\psi''_j$, where $\psi_j$
denotes the wave function at the $j$th edge. Since early times it
has  been known that a vertex joining $n$ graph edges can be
characterized by $n^2$ real parameters \cite{ES89} characterizing
the boundary condition at the vertex. We use the symbol $\Psi(0)$
for the column vector of the boundary values at the vertex
(identified conventionally with the origin of the coordinates),
and analogously $\Psi'(0)$ for the vector of the derivatives,
taken all in the outgoing direction.

The boundary conditions have to be chosen to make the Hamiltonian
self-adjoint, or in physical terms, to ensure conservation of the
probability current at the vertex. A general form of such a
coupling was found in \cite{KS99}. It is described by a pair of
$n\times n$ matrices $A,B$ such that $\mathrm{rank\,}(A,B)=n$ and
$AB^*$ is self-adjoint; the boundary values have to satisfy the
conditions
 % ------------- %
 \begin{equation} \label{KS bc}
 A\Psi(0)+B\Psi'(0)=0\,.
 \end{equation}
 % ------------- %
They have an advantage in comparison to earlier parameterizations
relating $\Psi(0)$ and $\Psi'(0)$ by a single matrix, because the
latter is typically singular for a subset of parameters, albeit a
zero-measure one.

On the other hand, the matrix pair in (\ref{KS bc}) is non-unique;
one would prefer to have a condition analogous to
$\psi(0)\cos\theta + \psi'(0)\sin\theta=0$ is case of a single
edge end. Such conditions exist, they were obtained independently
in \cite{FT00, CFT01} for a generalized point interaction, $n=2$,
and in \cite{Ha00} for any $n\ge 1$. It is easy to derive them:
the self-adjointness requires vanishing of the boundary form,
$\sum_{j=1}^n (\bar\psi_j \psi'_j - \bar\psi'_j \psi_j)(0)=0$,
which occurs iff the norms $\|\Psi(0)\pm
i\ell\Psi'(0)\|_{\mathbb{C}^n}$ with a fixed nonzero $\ell$
coincide, so the two vectors must be related by an $n\times n$
unitary matrix. The length parameter is not important because
matrices corresponding to two different values are related by
 % ------------- %
 \begin{equation} \label{choice l}
 U' = \frac{(\ell+\ell')U +\ell-\ell'}{(\ell-\ell')U
 +\ell+\ell'}\,.
 \end{equation}
 % ------------- %
Thus we set $\ell=1$, which means a choice of the length scale,
and put
 % ------------- %
 \begin{equation} \label{AB through U}
 A=U-I\,,\quad B=i(U+I)\,;
 \end{equation}
 % ------------- %
the edges are obviously fully decoupled at the vertex iff $U$ is
diagonal. It is easy to check that any such pair satisfies the
above quoted requirements from \cite{KS99}. Conversely, to any
$A,B$ with these properties there is a $U\in U(n)$ and an
invertible $C$ such that $U=C(A-iB)$. Indeed, such a $U$ must
satisfy $UU^*= C(BB^*+ AA^*)C^*$ since $AB^*=BA^*$ by assumption.
The matrix $BB^*+ AA^*$ is strictly positive because its null
space is
 % ------------- %
 \begin{equation} \label{kernel}
 \mathrm{ker\,}A^* \cap \mathrm{ker\,}B^* =
 (\mathrm{ran\,}A)^\perp \cap (\mathrm{ran\,}B)^\perp =
 (\mathrm{ran\,}A \cup \,\mathrm{ran\,}B)^\perp = \{0\}\,.
 \end{equation}
 % ------------- %
In particular, it is Hermitean so $C:= (BB^*+ AA^*)^{-1/2}$ makes
sense, it is Hermitean and invertible.

The parametrization (\ref{AB through U}) simplifies various
previous results. For instance, the eigenspace of $U$ with
eigenvalue $-1$ gives the projection $P=P_1$ in \cite{Ku04} which
makes Lemma~4 and the following claims of this paper rather
transparent. Likewise, the on-shell scattering matrix for a star
graph of $n$ halflines with the considered coupling equals
 % ------------- %
 \begin{equation} \label{on-shell S}
 S_U(k) = \frac{(k-1)I+(k+1)U}{(k+1)I+(k-1)U}\,,
 \end{equation}
 % ------------- %
which makes a discussion of its properties simpler than in Sec.~2
of \cite{KS99}.

To give an example of the parametrization (\ref{AB through U}),
denote by $\JJ$ the $n\times n$ matrix whose all entries are equal
to one. It is a straightforward exercise to check that $U= {2\over
n+i\alpha}\JJ-I$ describes the standard $\delta$ \emph{coupling},
 % ------------- %
 \begin{equation} \label{delta}
 \psi_j(0)=\psi_k(0)=:\psi(0)\,,\; j,k=1,\dots,n\,, \quad \sum_{j=1}^n
 \psi'_j(0) = \alpha \psi(0)
 \end{equation}
 % ------------- %
with $\alpha\in\R$; the case $\alpha=0$ corresponds to the ``free
motion'' at the vertex, so-called Kirchhoff boundary conditions,
while $\alpha=\infty$ gives $U=-I$, the full Dirichlet decoupling.
In a similar way, $U= I-{2\over n-i\beta}\JJ$ describes the
singular counterpart, so-called $\delta'_s$ \emph{coupling}
\cite{Ex95, Ex96a},
 % ------------- %
 \begin{equation} \label{delta'_s}
 \psi'_j(0)=\psi'_k(0)=:\psi'(0)\,,\; j,k=1,\dots,n\,, \quad \sum_{j=1}^n
 \psi_j(0) = \beta \psi'(0)
 \end{equation}
 % ------------- %
with $\beta\in\R$; for $\beta=\infty$ we get $U=I$, the full
Neumann decoupling.

Let us mention another ``dual'' pair of vertex couplings in which
the wave functions exhibit permutation symmetry. The more regular
one of these is the ``permuted'' $\delta$, or $\delta_p$
\emph{coupling}, given by the boundary conditions
 % ------------- %
 \begin{equation} \label{delta_p}
 \sum_{j=1}^n \psi_j(0)=0\,,\quad
 \psi'_j(0)-\psi'_k(0) =
 {\alpha\over n}
 (\psi_j(0)-\psi_k(0))\,,\; j,k=1,\dots,n\,,
 \end{equation}
 % ------------- %
with $\alpha\in\R$. It generalizes the $\delta_s$ interaction of
\cite{TFC01} and one can check easily that the corresponding
matrix equals $U= {n-i\alpha\over n+i\alpha}I-
\frac{2}{n+i\alpha}\JJ$. Its counterpart is the so-called
$\delta'$ \emph{coupling} \cite{Ex95, Ex96a},
 % ------------- %
 \begin{equation} \label{delta'}
 \sum_{j=1}^n \psi'_j(0)=0\,,\quad
 \psi_j(0)-\psi_k(0) =
 {\beta\over n}
 (\psi'_j(0)-\psi'_k(0))\,,\; j,k=1,\dots,n\,,
 \end{equation}
 % ------------- %
with $\beta\in\R$, which corresponds to $U= -{n+i\beta\over
n-i\beta}I+ \frac{2}{n-i\beta} \JJ$. The infinite values of the
parameters refer again to the Dirichlet and Neumann decoupling of
the graph edges, respectively.

Note that in these four examples, the connection condition at the
origin is totally symmetric with respect to the interchange of
edge indices.  Consequently, their $U$ are constructed from
symmetric matrices $I$ and ${\cal J}$.

If one wants to continue analysis of such graphs, the first
question to be answered is about the physical meaning and possible
use of the whole family of such general couplings. What concerns
the second part, a recent inspiration comes from the domain of
quantum computing, where the generalized point interactions
parameterized by elements of the group $U(2)$ have been proposed
as an alternative realization of a qubit \cite{CFT04}; an
extension to higher degree vertices opens, of course, interesting
possibilities. To make use of them, however, one has to understand
whether there is a meaningful way to ``construct'' vertices with
different couplings.

The currently available results suggest that this goal cannot be
achieved in a purely geometrical way, by squeezing a system of
branching tubes with the same topology as the graph. Several such
approximations was analyzed recently \cite{KZ01, RS01, Sa01,
EP03}; they all lead either to trivial (Kirchhoff) boundary
conditions, or to graphs having an extended state Hilbert space
with extra dimensions due to the vertices. Their common feature
was that the transverse ground state was a constant function.
Hence a nontrivial results might be obtained through tubes with
Dirichlet boundaries, but this problem is open for a long time and
notoriously difficult.

Approximations using potentials scaled in the usual way, i.e.
preserving their integrals, do yield nontrivial results
\cite{Ex96b} but only for couplings with wave functions continuous
at the junction, which is just the family~(\ref{delta}). This is
not sufficient and more singular coupling need other means. Our
main aim here is to explore a natural alternative with
approximating interactions scaled in a \emph{nonlinear} way as a
generalization of the procedure proposed in \cite{CS98a, CS98b}
and analyzed from the viewpoint of the convergence topology in
\cite{AN00, ENZ01}. To keep things simple we will analyze here the
$\delta'_s$ and $\delta'$ couplings leaving the general case to a
subsequent paper.

Consider first the Hamiltonian $H_\beta$ on the graph $\Gamma$
consisting on $n$ halflines coupled at a single vertex by the
conditions (\ref{delta'_s}). Consider further the same graph with
additional vertices of degree two at each arm, all at the same
distance $a>0$ from the common junction. The approximating family
will be constructed as follows. The operators act, of course, as
$\psi_j \mapsto -\psi''_j$ at each arm; the wave functions satisfy
the $\delta$ \emph{conditions} (\ref{delta}) at the central vertex
with a coupling parameter $\alpha=b$, to be specified later, and
another $\delta$ \emph{coupling} (\ref{delta}), this time with the
parameter $c$, at each of the additional vertices -- see Fig~1. We
call such an operator $H^{b,c}(a)$.

 % ------------- %
 \begin{figure}
 \setlength\unitlength{1mm}
 \begin{picture}(95,0)(38,95)
 \thicklines
 \put(70,40){\line(-1,-1){10}}
 \put(70,40){\line(2,-1){12}}
 \put(70,40){\line(1,1){15}}
 \put(70,40){\line(-1,1){10}}
 \put(70,40){\circle*{.8}}
 \put(140,40){\line(-1,-1){10}}
 \put(140,40){\line(2,-1){12}}
 \put(140,40){\line(1,1){15}}
 \put(140,40){\line(-1,1){10}}
 \put(140,40){\circle{.8}}
 \put(100.5,39){\mbox{\Large $\longrightarrow$}}
 \put(100,35){\mbox{$a\to 0$}}
 \put(138.5,42.5){\mbox{$\beta$}}
 \put(68.3,42){\mbox{$a$}}
 \put(70.6,21){\mbox{$b(a)$}}
 \put(73.8,30.8){\mbox{$c(a)$}}
 \put(121,52){\mbox{$H_\beta$}}
 \put(50,52){\mbox{$H^{b,c}$}}
 \thinlines
 \put(70,40){\line(0,-1){19}}
 \put(66.5,36.5){\line(0,-1){8}}
 \put(73.2,38.4){\line(0,-1){8}}
 \put(72.4,42.4){\line(0,-1){8}}
 \put(67,42.9){\line(0,-1){8}}
 \put(44,25){\line(1,0){40}}
 \put(44,25){\line(1,2){16}}
 \put(84,25){\line(1,2){16}}
 \put(60,57){\line(1,0){40}}
 \put(114,25){\line(1,0){40}}
 \put(114,25){\line(1,2){16}}
 \put(154,25){\line(1,2){16}}
 \put(130,57){\line(1,0){40}}
 \end{picture}
 \vspace{75mm}
 \caption{Scheme of the approximation. For simplicity the graph
 is featured as planar; the vertical bars denote the $\delta$
 coupling strength.}
 \end{figure}
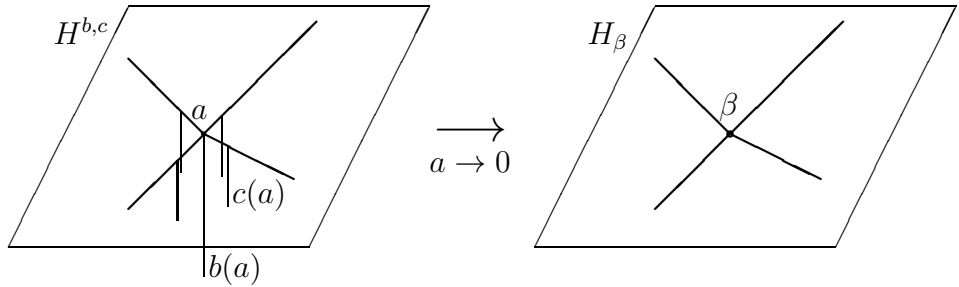
 % ------------- %

The crucial feature that allows us to simplify the treatment in
the present situation is a symmetry. Each of the Hamiltonians
$H_\beta$ and $H^{b,c}(a)$ decomposes into a nontrivial part which
acts on the one-dimensional subspace of $\HH= \bigoplus_{j=1}^n
L^2(\R_+)$ consisting of functions symmetric with respect to
permutations, $\psi_j(x)=\psi_k(x)$ for all $j,k$, and the
$(n\!-\!1)$-dimensional part corresponding to Dirichlet and
Neumann condition at the central vertex for the $\delta$ and
$\delta'_s$ coupling, respectively. Notice that the matrices $U$
corresponding to these coupling have each one simple eigenvalue
and another one equal to $\mp1$, respectively, of multiplicity
$n-1$.

To see what the choice of the effective coupling constants $b,c$
should be, let us first modify to our problem the argument of
\cite{CS98a}. As we have said, in the nontrivial sector all the
functions are the same, so we may drop the arm index. The boundary
values at $x=0$ and $x=a$ are related by
 % ------------- %
 \begin{eqnarray}
 \psi(a) = \psi(0)+a\psi'(0)+\OO(a^2)\,, &&\hspace{-1em}
 \psi'(a-)=\psi'(0+)+\OO(a)\,, \label{exp delta} \\
 \psi'(a+) = \psi'(a-)+c\psi(a)\,, &&\hspace{-1em}
 \psi'(0+)=b\psi(0)\,. \label{bc delta}
 \end{eqnarray}
 % ------------- %
Eliminating $\psi(0)$ and $\psi'(0+)$ from here, we get in the
leading order the relation $B(a)\psi(a)= \psi'(a+)$, where
 % ------------- %
 \begin{equation} \label{B}
 B(a):= c + \frac{b}{1+ab}\,;
 \end{equation}
 % ------------- %
hence the needed limit, $\beta\psi'(0+)=n\psi(0)$, is achieved as
$a\to 0+$ if we choose
 % ------------- %
 \begin{equation} \label{bc}
 b(a):= -\frac{\beta}{na^2}\,, \quad c(a):= -\frac{1}{a}\,.
 \end{equation}
 % ------------- %
In the orthogonal complement to the permutation-symmetric subspace
one we can again drop the index, because the operators act in the
same way on all the linear combinations of $\sum_{j=1}^n
d_j\psi_j(x)$ which we can choose as the basis here, i.e. those
satisfying $\sum_{j=1}^n d_j=0$. The second one of the conditions
(\ref{bc delta}) is now replaced by $\psi(0)=0$. Eliminating then
the boundary values at $x=0$ we get in the leading order the
relation $\psi'(a+)= (c+a^{-1})\psi(a)+\OO(a)$. The right-hand
side vanishes with the parameter choice (\ref{bc}), giving Neumann
condition, $\psi'(0+)=0$, in the limit.

Now we can state and prove our main result. \vspace{.5em}

 % ------------- %
 \noindent \textbf{Theorem 1} \textit{
 $H^{b,c}(a)\to H_\beta$ as $a\to 0+$ in the norm-resolvent
 sense provided the coupling constants $b,c$ are
 chosen in correspondence with (\ref{bc}).}
 \vspace{.5em}
 % ------------- %

\noindent \textsl{Proof:} By the same symmetry argument as above
we can again reduce the problem to investigation of a pair of
halfline problems. Let us start with the one having Dirichlet
condition at the origin, so the free Green's function at energy
$k^2$ is
 % ------------- %
 \begin{equation} \label{D Green}
 G_k(x,y)= \frac{\sin kx_<}{k}\, \e^{ikx_>}\,,
 \end{equation}
 % ------------- %
where as usual $x_<$ is the smaller one of the values $x,y$ and
vice versa. The Green's function of the operator with the $\delta$
interaction at $x=a$ is obtained easily by Krein's formula
\cite[Appendix A]{AGHH}
 % ------------- %
 \begin{equation} \label{D Krein}
 G^c_k(x,y)= G_k(x,y)+ \frac{G_k(x,a)G_k(a,y)}{-c^{-1}-G_k(a,a)} \,.
 \end{equation}
 % ------------- %
On the other hand, the Green's function referring to Neumann
boundary is
% 16
 % ------------- %
 \begin{equation} \label{D Neumann}
 G^N_k(x,y)= \frac{\cos kx_<}{k}\, \e^{ikx_>}\,;
 \end{equation}
 % ------------- %
our aim is to show that the last two converge to each other for
some $k^2\in\C$. It is convenient to choose $k=i\kappa$ with
$\kappa>0$; we will see below that the denominator of the last
term at the right-hand side of (\ref{D Krein}) is then nonzero for
$a$ small enough. Since the functions involved are uniformly
bounded around zero, it is sufficient to compute the difference in
the case when neither of the arguments is smaller than $a$. For
the sake of definiteness suppose that $a\le x\le y$; then (\ref{D
Krein}) and (\ref{D Neumann}) give
 % ------------- %
 \begin{equation} \label{D diff}
 G^c_{i\kappa}(x,y)- G^N_{i\kappa}(x,y) = \frac{\e^{-\kappa x}
 \e^{-\kappa y}}{\kappa} \left[ -1+ \frac{\sinh^2\kappa a}
 {-\kappa c^{-1}- \e^{-\kappa x} \sinh^2\kappa a } \right]\,.
 \end{equation}
 % ------------- %
If $c=-a^{-1}$ the last term behaves as $1+\OO(a)$ for $a\to 0+$,
so
 % ------------- %
 \begin{equation} \label{D limit}
 \lim_{a\to 0+} G^c_{i\kappa}(x,y)= G^N_{i\kappa}(x,y)
 \end{equation}
 % ------------- %
holds for all $x,y>0$.

Consider next the case with the $\delta$ coupling at the origin
using the same parameter values, namely $k=i\kappa$ and $a\le x\le
y$. We are interested in the following two Green's functions,
 % ------------- %
 \begin{eqnarray}
 && G^b_{i\kappa}(x,y) = \frac{\e^{-\kappa y}}{\kappa
 (b+\kappa)}\,
 (b\sinh\kappa x + \kappa\cosh\kappa x)\,,
 \label{delta free Green} \\
 && G^\beta_{i\kappa}(x,y) = \frac{\e^{-\kappa y}}{\kappa
 (n+\beta\kappa)}\, (n\sinh\kappa x + \beta\kappa\cosh\kappa x)\,,
 \label{beta Green}
 \end{eqnarray}
 % ------------- %
which replace (\ref{D Green}) and (\ref{D Neumann}), respectively,
in the present case. The first of them determines the full
approximating Green's function by Krein's formula,
 % ------------- %
 \begin{equation} \label{delta Krein}
 G^{b,c}_k(x,y)= G^b_k(x,y)+ \frac{G^b_k(x,a)G^b_k(a,y}
 {-c^{-1}-G^b_k(a,a)} \,.
 \end{equation}
 % ------------- %
Using the relations (\ref{bc}) we express the difference
 % ------------- %
 \begin{eqnarray}
 \lefteqn{G^{b,c}_{i\kappa}(x,y)- G^\beta_{i\kappa}(x,y) = \frac{
 \e^{-\kappa y}}{\kappa}\, \Bigg[ \frac{b\sinh\kappa x +
 \kappa\cosh\kappa x}{b+\kappa} } \nonumber \\ &&
 + \frac{\frac{\e^{-\kappa x}}{(b+\kappa)^2} (b\sinh\kappa x +
 \kappa\cosh\kappa x)^2}{\kappa a - \frac{\e^{-\kappa a}}{b+\kappa}
 (b\sinh\kappa x + \kappa\cosh\kappa x)} -
 \frac{n\sinh\kappa x + \beta\kappa\cosh\kappa x}{n+\beta\kappa}
 \Bigg] \label{delta diff}
 \end{eqnarray}
 % ------------- %
The first term in the bracket tends to $\sinh \kappa x$ as $a\to
0+$, while the third one is independent of $a$, so their sum in
the limit gives
 % ------------- %
 \begin{equation} \label{13}
 - \,\frac{\beta\kappa\, \e^{-\kappa x}}{n+\beta\kappa}\,.
 \end{equation}
 % ------------- %
Next we take the middle term without the factor $\e^{-\kappa x}$
and expand the numerator and denominator to the second power in
$a$; this gives its limit which differs just by the sign from
(\ref{13}), in other words
 % ------------- %
 \begin{equation} \label{delta limit}
 \lim_{a\to 0+} G^{b,c}_{i\kappa}(x,y)= G^\beta_{i\kappa}(x,y)
 \end{equation}
 % ------------- %
holds again for all $x,y>0$. To conclude the proof we have just to
realize that as functions of $x,y$ the differences (\ref{D diff})
and (\ref{delta diff}) decay exponentially, so the corresponding
resolvent differences converge to zero even in the Hilbert-Schmidt
norm.$\;\Box$ \vspace{.5em}

Let us add that the proven result opens way to approximation of
$H_\beta$ by Hamiltonians with more regular interactions. We have
mentioned that the central $\delta$ in $H^{b,c}(a)$ can be
approximated by a family of potentials scaled in the usual way,
the same is true for the $\delta$ interactions at the points
$x_j=a$. As in the related problem discussed in \cite{ENZ01}, the
question then is how fast have these approximating potentials to
shrink with respect to $a$.

Consider finally the case of $\delta'$ \emph{coupling}, i.e. the
Hamiltonian $\tilde H_\beta$ on our star graph with the boundary
conditions (\ref{delta'}) at the vertex. The approximating family
will be constructed in a similar way as above; the difference is
that now the wave functions will satisfy the $\delta_p$
\emph{conditions} (\ref{delta_p}) at the central vertex with a
coupling parameter $\alpha=b$, to be specified. The rest is the
same, there is another $\delta$ \emph{coupling} (\ref{delta}) with
the parameter called again $c$ at each of the additional vertices;
we denote such an operator $\tilde H^{b,c}(a)$.

To realize that this problem can be again reduced to a
one-dimensional analysis, denote $\varepsilon:= \e^{2\pi i/n}$ and
introduce
 % ------------- %
 \begin{equation} \label{decomp}
 \GG_r:= \left\{ (\psi(x), \varepsilon^r \psi(x), \dots,
 \varepsilon^{r(n-1)} \psi(x))\,: \: \psi\in L^2(\R_+) \right\}\,.
 \end{equation}
 % ------------- %
The graph state Hilbert space can be written as $\HH=
\bigoplus_{r=0}^{n-1} \GG_r$. Indeed, any vector of $\HH$ is a
unique linear combination of the vectors from $\GG_r$, because the
determinant of the corresponding linear system is the Vandermond
determinant of $1,\varepsilon,\dots, \varepsilon^{n-1}$, which is
nonzero because the latter are mutually different. It is
straightforward to see that the subspaces $\GG_r$ are invariant
under the Hamiltonians in question; the $\delta_p$ and $\delta'$
couplings acts ``trivially'' at the origin corresponding to
Dirichlet and Neumann condition, respectively, while on each of
the subspaces $\GG_r,\: r=1,\dots,n$, these boundary conditions
are replaced by $\psi'(0)= \frac{\alpha}{n}\psi(0)$ and $\psi(0)=
\frac{\beta}{n}\psi'(0)$. Thus we can choose
 % ------------- %
 \begin{equation} \label{bc_p}
 b(a):= -\frac{\beta}{a^2}\,, \quad c(a):= -\frac{1}{a}\,,
 \end{equation}
 % ------------- %
and repeat the above considerations, arriving at the following
conclusion. \vspace{.5em}

 % ------------- %
 \noindent \textbf{Theorem 2} \textit{
 $\tilde H^{b,c}(a)\to \tilde H_\beta$ holds in the norm-resolvent \
 sense as $a\to 0+$  if the coupling constant families $b,c$ are
 given by (\ref{bc_p}).}
 \vspace{.5em}
 % ------------- %

\subsection*{Acknowledgments}

P.E. appreciates the hospitality extended to him at the Kochi
University of Technology where a part of this work was done. The
research has been partially supported by Czech Ministry of
Education and ASCR within the projects ME482 and K1010104.

\end{document}